\RequirePackage[l2tabu, orthodox]{nag}
\RequirePackage{fix-cm}
  \documentclass[12pt,onecolumn,letter,draftclsnofoot]{IEEEtran}
\IEEEoverridecommandlockouts
\usepackage{etoolbox}
\patchcmd{\section}{\centering}{}{}{}

\usepackage{mdframed}
\usepackage[makeroom]{cancel}

\usepackage{graphicx} 

\usepackage[cmex10]{amsmath} 
\usepackage{amssymb}
\usepackage{amsfonts}
\usepackage{mathrsfs} 

\usepackage{cuted}

 \usepackage{subcaption}

\usepackage[noadjust]{cite}

\usepackage[x11names]{xcolor}   

\usepackage{soul}
\usepackage[T1]{fontenc}
\usepackage{charter}
\usepackage{environ}
\usepackage{tikz}
\usetikzlibrary{calc,matrix}

\usepackage[section]{placeins} 
\usepackage{stackengine}

\usepackage{enumitem}
\setlist[enumerate]{label=\roman*} 

\usepackage{microtype}

\usepackage[colorlinks]{hyperref}
\hypersetup{%
colorlinks = true,
citecolor=Green4,
linkcolor  = blue
}

\pagebreak

\ifCLASSINFOpdf
\else
\fi

\begin{document}

\title{Estimation in the Gaussian Multiplex Channel}

\author{\IEEEauthorblockN{Daniel R. Fuhrmann,~\IEEEmembership{Fellow,~IEEE}} \\
\IEEEauthorblockA{Department of Applied Computing,\\
Michigan Technological University\\ 
Houghton, MI 49931, USA \\
fuhrmann@mtu.edu}
\and \\
\IEEEauthorblockN{Muhammad Fahad,~\IEEEmembership{Member,~IEEE}} \\
\IEEEauthorblockA{mfahad@mtu.edu}
}

\maketitle
\begin{abstract} \label{abstract}
An abstraction for multisensor communication termed the Gaussian
Multiplex Channel is presented and analyzed.  In this model, the sensor
outputs can be added together in any combination through a network of
switches, and the combinations can be changed arbitrarily during the
observation interval.  The sensor output sums are observed in additive
Gaussian noise.  Using a mean square error cost function and a
constraint on the total observation time, an optimal set of combinations
(switch positions) and observation times is determined.  The solution
exhibits high complexity (number of different combinations) even for
moderate numbers of sensors.  It is then shown that there exists an
alternative solution based on Hadamard designs, which achieves the same
minimizing MSE cost function and only requires a number of combinations
equal to the number of sensors.
\end{abstract}
\section{Introduction} \label{vectorchannel}
This paper considers one version of a problem in which there are many sensors, one communication channel, and a limited amount of time to acquire data and perform statistical inference \cite{hero2007foundations}. The time aspect of the problem may be due to operational requirements to make decisions and take actions, and may be due to the dynamic nature of the parameters under investigation, which may be treated as constant for times.  

We present a problem as a "thought experiment" involving multiple sources and switches and a desire to estimate certain parameters as accurately as possible given the system constraints. While our proposed system may not match a given system exactly, it is hoped that the results obtained may be useful in applications that share some of the attributions for the abstraction.

This work is inspired by another problem in the Bayesian framework where only the mean vector of a multivariate Gaussian is affected by the sensing times under a given total available time constraint while the covariance matrix is assumed as an identity matrix is discussed in \cite{paper2} and \cite{paper3}. 

\section{Problem Statement and Example Solutions}

\begin{figure}[t] 
\centering
\includegraphics[width=0.8\linewidth]{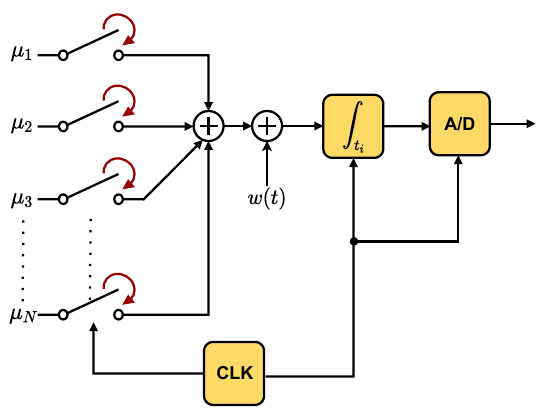}  
\caption{The Gaussian Multiplex Channel}
\label{f0}
\end{figure}

A block diagram for the abstraction we call the Gaussian Multiplex Channel is shown in Fig.\ref{f0}.  There $N$ parameters, $\mu_1, \cdots \mu_N$ presumably the output of $N$ different sensors.  These parameters can be added together in any combination through a network of switches as shown in the figure. Switches can be opened and closed throughout the data acquisition process.

Added to the sum of the parameters, as seen through the switches, is a white noise process $w(t)$ with the property that $\int_0^T w(t) \,dt \sim \mathcal{N}(0,\sigma^2 T)$. Without loss of generality, take $\sigma^2=1$.

The total time available to estimate $\mu_1 \cdots \mu_N$ is $T$ seconds. Again, without loss of generality take $T=N$ so that we have $1$ sec per parameter.  The symbol $T$ will later be used to denote a certain diagonal matrix, so $N$ will be used to denote both the number of parameters and the time available to estimate them.

The available time is divided into $M$ segments $(M \ge N)$. $t_1 \cdots t_M \: \operatorname{s.t.} \: \sum_{i=1}^{M} t_i=N$. In each time segment, there is a unique configuration of switch positions given by a binary vector $b_i^{\top}=[b_i(1) \cdots \cdots b_i(N)]$, where $b_i = 1$ denotes a switch being closed and $b_i=0$ denotes that switch being open.  At the beginning of each time segment, the integrator is reset, and at the end the integrator is sampled to produce a random variable $X_i$. From this model, we have 
\begin{IEEEeqnarray*}{rCl}
X_i&=&t_i (b_i^{\top} \mu) + \int_0^{t_i} w(t) \,dt \label{2.1} \IEEEyesnumber \\
X_i & \sim & \mathcal{N} \Big( t_i(b_i^{\top}  \mu),t_i \Big) 
\label{E2} \IEEEyesnumber
\end{IEEEeqnarray*} 	
Defining the observation vector as $X=[X_1 \cdots \cdots X_M]^{\top}$, where have the matrix-vector model
\begin{IEEEeqnarray*}{rCl}
X&\sim & \mathcal{N} \big( T B \mu,T \big),  \label{E3} \IEEEyesnumber
\end{IEEEeqnarray*}
where $B=\begin{bmatrix}
b_1^{\top}\\
b_2^{\top} \\
\vdots \\
b_M^{\top}
\end{bmatrix}$ and $T=\begin{bmatrix}
t_1 & \cdots & 0\\
& \ddots & \\
0 & \cdots & t_M
\end{bmatrix}.\\$

$B$ is an $M \times N$ matrix whose rows are binary vectors describing the switch positions and $T$ is a diagonal matrix of $\operatorname{diag}(t_1 \cdots t_M) \: \operatorname{s.t.} \: \operatorname{Tr} T=N$.

Given the independent increments nature of $w(t)$,
it suffices to make only one observation for each unique vector of switch positions. Excluding the all-$0$s row (when all switches would be open), there are $2^N-1$ rows of $B$, $M=2^N-1$ times $t_1 \cdots t_M$. If any of the $t_i$'s are $0$, we can delete there corresponding row of $B$.

Given $B$ and $T$ we can easily find the $\mu_{\text{ML}}$ estimate of $\mu$ which we shall do shortly.  As we shall see $\hat{\mu}_{\text{ML}}\sim \mathcal{N}(\mu,R)$ where $R$ is a function of $B$ and $T$. Thus the real problem we address is selection of switch positions $(B)$ and times $(T)$ to minimize $\operatorname{Tr} R$.

Before proceeding with the analysis, a couple of comments about the model.  In essence, what we are doing is exploring a trade-off between signal-to-noise ratio (SNR) and ambiguity in the measurements.  Any measurement equal to the sum of multiple parameters inherently has an ambiguity about which parameters contribute what to the sum.  On the other hand, looking at multiple signals simultaneously increases the SNR of any given measurement.  It should be noted that the fact there is only one noise source, $w(t)$, which is associated with the data acquisition and communication channel, is a key feature of the model and the analysis.  If each sensor brings with it its own additive noise, then "all bets are off" as adding them together does nothing to increase the SNR - in which case one cannot do better that observing each parameter in isolation.

It should also be noted that we are not really putting any bandwidth constraints on the channel, and indeed if $M$ is much greater than $N$ then one may be in a position of having to transmit a huge amount of data in a short period of time.  This possibility will appear to exist in the analysis of Section \ref{S4}.  However, in the end we shall find a way to achieve optimal performance with only $N$ measurements, which may keep the communication requirements reasonable. 

\section{ML Estimation of \texorpdfstring{$\mu$} and MSE}

We begin with standard maximum-likelihood estimation of the vector $\mu$ and the resulting mean square error (MSE).  The probability density function for the observation is
\begin{IEEEeqnarray*}{rCl}
f_X(x)&=& \pi^{-\frac{M}{2}}(\operatorname{det} T)^{-\frac{1}{2}} e^{-\frac{(x-TB\mu)^{\top} T^{-1}(x-TB\mu)}{2}}\label{E4} \IEEEyesnumber
\end{IEEEeqnarray*}
The log-likelihood with constant terms removed is
\begin{IEEEeqnarray*}{rCl}
L(\mu)&=&(x-TB\mu)^{\top} T^{-1}(x-TB\mu) \\
&=& x^{\top} T^{-1} x -2 x^{\top}T^{-1}TB \mu +\mu^{\top}B^{\top}T T^{-1}T B \mu \\
&=& x^{\top}T^{-1}x-2x^{\top}B \mu+\mu^{\top}B^{\top}T B \mu
\label{E5} \IEEEyesnumber
\end{IEEEeqnarray*}
The gradient of $L$ w.r.t $\mu$ is
\begin{IEEEeqnarray*}{rCl}
\nabla L &=& -2 B^{\top}x+2 B^{\top}TB \mu .
\label{E7} \IEEEyesnumber
\end{IEEEeqnarray*}
Setting this to $0$ we get
\begin{IEEEeqnarray}{rCl}
B^{\top}TB\mu=B^{\top}x
\label{E8} 
\end{IEEEeqnarray}
or \begin{IEEEeqnarray}{rCl}
\hat{\mu}_{\text{ML}}&=&(B^{\top}TB)^{-1}B^{\top}x .
\label{E08} 
\end{IEEEeqnarray}
The maximum-likelihood estimator can be viewed as a random variable (a function of a random variable $X$), i.e.
\begin{IEEEeqnarray}{rCl}
\hat{\mu}_{\text{ML}}&=&(B^{\top}TB)^{-1}B^{\top}X .
\label{E9} 
\end{IEEEeqnarray}
Based on the statistical model for $X$, given in (3) above, 
we have that $\mathbb{E}[\hat{\mu}_{\text{ML}}]= \allowbreak (B^{\top}TB)^{-1}\allowbreak B^{\top}TB\mu=\mu$ (the estimator is unbiased) and 
\begin{IEEEeqnarray}{rCl}
\operatorname{cov}(\hat{\mu}_{\text{ML}})&=&(B^{\top}TB)^{-1}B^{\top}TB(B^{\top}TB)^{-1} \\
&=&(B^{\top}TB)^{-1} 
\label{E11} 
\end{IEEEeqnarray}

The analysis above shows that it is possible to control the estimator performance through control of the switch positions and the amount of time spent in each configuration.  Hence, our aim is to minimize $\operatorname{Tr} (B^{\top}TB)^{-1} $ subject to the constraint that $B$ is a binary $0-1$ matrix with $M$ unique rows and $T$ is a diagonal matrix with $\operatorname{Tr} T=N$.  The remainder of this paper deals with this optimization problem.

For notational convenience we introduce
$$C=B^{\top}TB$$
It is well-known for linear Gaussian estimation problem like this that $C$ is the Fisher Information Matrix (FIM) and $\operatorname{Tr} C^{-1}$ is the Cramer-Rao bound. The ML estimator is efficient in this case.  Note that he MSE does not depend on the parameter $\mu$. The estimator $\hat{\mu}_{\text{ML}}$ is unbiased and all we are controlling through the choice of $B$ and $T$ is the distribution of $\hat{\mu}_{\text{ML}}$ about its mean $\mu$.  

One other notation:  define $J=11 ^{\top}$, the all-ones matrix.
The dimensions of $J$ will normally be easily determined from context; if necessary we can write $J_{N}=$ the $N \times N$ all-ones matrix.

\section{Examples}

Before proceeding to the full analysis of the optimization, we first give a few examples of what one might do based on intuition.

\subsection{Example $1$.} \label{L1}
As a simple example we look at each target in isolation for $1$ second each. Each vector of switch positions has a single $1$, so that $B=I$ and $T=I$ as well. Then

$ C=(I)(I)(I)=I $ and $\operatorname{Tr} C^{-1}=N$.
Thus, we observe each target for $1$ second and the corresponding entry in $\hat{\mu}_{\text{ML}}$ has variance $1$. Each entry in $\hat{\mu}_{\text{ML}}$ is independent.

\subsection{Example $2.$} \label{L2} As a second example, suppose that again $M=N$ but that each row of $B$ corresponding to $N-1$ switches being \emph{closed} instead of open. In this case, we significantly increase the SNR in each observation but since many parameters are being added there is \emph{ambiguity} in the way that each target contributes to each observation.

We have that $B=\begin{bmatrix}
1 & 1 & 1 & 1 \cdots \cdots & 1 & 1 & 0\\
1 & 1 & 1 & 1 \cdots \cdots & 1 & 0 & 1\\
\cdots  & \cdots & \cdots  & \cdots \cdots& \cdots & \cdots & \cdots\\
0  & 1 & 1 & 1 \cdots \cdots& 1 & 1 & 1\\
\end{bmatrix}$

Take $T=I$ as before. Then
$C=B^{\top}B=\begin{bmatrix}
N-1 & N-2 & \cdots & N-2\\
N-2 & \ddots &  & \\
\vdots &  & \cdots & N-1\\
\end{bmatrix}\\$

We will see the structure again many times, a symmetric matrix with one value on the diagonal and a smaller value on all the off-diagonals. We can write this as
\begin{IEEEeqnarray}{rCl}
\big((N-1)-(N-2)\big)I+(N-2)J&=&I+(N-2)J
\label{E12} 
\end{IEEEeqnarray}

The eigenvalues of $C$ are $1,1,1 \cdots 1,N-2+1$ or $1,1,1, \cdots 1,N-1$.

The eigenvalues of $C^{-1}$ are $1,1,1 \cdots 1, \frac{1}{N-1}$.

Thus 
\begin{subequations}
\begin{IEEEeqnarray}{rCl}
\operatorname{Tr} C^{-1}&=&\sum_{i=1}^N (\lambda_i)^{-1} \\
&=&1+1+1+ \cdots \cdots 1+\frac{1}{N-1} \\
&=& N-1+\frac{1}{N-1} \\
&=& \frac{(N-1)^2+1}{N-1}= \frac{N^2-2N+2}{N-1}
\label{E13} 
\end{IEEEeqnarray}
\end{subequations}
and for large $N$, $\operatorname{Tr} C^{-1} \approx N-1$.

Thus this method provides a very slight improvement in MSE over example $1.$ given in subsection $\ref{L1}$.  These first two examples are at the extremes of the ambiguity/SNR tradeoff: the first example has low SNR and no ambiguity, whereas the second has high SNR and high ambiguity.

\subsection{Example $3.$} \label{L3} 
Let us return to the case where one observes the targets one at a time but with unequal times $t_1 \cdots t_N$, still with the constraint $\sum_{i=1}^{N} t_i=\operatorname{Tr} T=N$.
Then $C=T$ and $\operatorname{Tr} C^{-1}=\sum_{i=1}^{N} \frac{1}{t_i}$.

The function $f(t)=\sum_{i=1}^{N} \frac{1}{t_i}$ is a symmetric Schur-convex function in $t$. Since $t \succ [1,1, \cdots 1]$ we know $f(t) \ge f(1)$ and thus $\operatorname{Tr} C^{-1}$ is minimized when the observation times are all equal. Thus there is no advantage to trying to mix up the times - equal times is the best.  This theme will reappear multiple times.

\subsection{Example $4.$} \label{L4} 
\begin{figure}[t] 
\centering
\includegraphics[width=\linewidth]{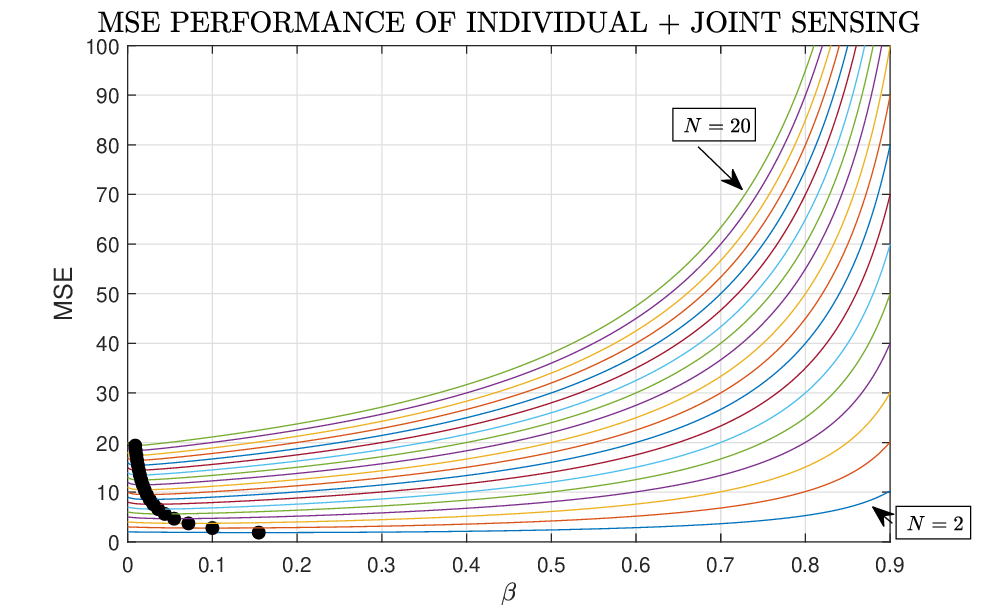}  
\caption{MSE performance of individual + joint sensing. 	}
\label{f1}
\end{figure}

As a final example before considering a more general approach, suppose we observe each target in isolation, for an equal amount of time, and then we reserve a little time at the end to look at all the targets at once, i.e., with all switches closed. In this case we have $M=N+1$ observations.
\begin{IEEEeqnarray}{rCl}
B=\begin{bmatrix}
I\\
1^{\top}
\end{bmatrix} \quad \text{and}  \quad T=\begin{bmatrix}
\alpha I  & \\
& N \beta
\end{bmatrix}.
\end{IEEEeqnarray}
$(\alpha+\beta=1), \operatorname{Tr} T= N$
\begin{IEEEeqnarray}{rCl}
C&=&B^{\top}TB=\begin{bmatrix}
I & 1\\

\end{bmatrix}\begin{bmatrix}
\alpha I & 0\\
0 & N \beta\\
\end{bmatrix}\begin{bmatrix}
I\\
1^{\top}
\end{bmatrix} \\
&=& \alpha I +N \beta J
\label{E14} 
\end{IEEEeqnarray}
The eigenvalues of $C$ are $\alpha,\alpha,\alpha, \cdots ,\alpha+N^2\beta $ and the eigenvalues of $C^{-1}$ are $ \allowbreak \frac{1}{\alpha},\frac{1}{\alpha},\frac{1}{\alpha}, \allowbreak \cdots \allowbreak \frac{1}{\alpha+N^2\beta}$.
\begin{IEEEeqnarray}{rCl}
\operatorname{Tr} C^{-1}=\frac{N}{\alpha}+\frac{1}{\alpha+N^2 \beta}
\end{IEEEeqnarray}
Substitute $\alpha=1-\beta$
\begin{IEEEeqnarray}{rCl}
\operatorname{Tr} C^{-1}=\frac{N}{1-\beta}+\frac{1}{1+(N^2-1) \beta}
\end{IEEEeqnarray}
This is minimized for values of $\beta$ which rapidly approach $0$ as $N$ increases.  See  Fig. \ref{f1}, which depicts the MSE as a function of $\beta$ for values of $N$ ranging for $2$ to $20$, and minimum of each curve shown as a black dot.  The advantage of the observation with all switches closed quickly becomes negligible as $N$ increases. The only real noteworthy case is $N=2, M=3$, which will turn out to be a special case of the general solution in Section \ref{L5}.

\section{Solution to Optimization Problem} \label{S4}

This section develops the switching strategy which leads to the minimum of the MSE.
\subsection{$k$ switches closed, for single $k$} \label{L5}
First consider a strategy in which in each time interval, we allow $k$ switches to be closed, where $1 \le k \le N-1$. The $k=1$ and $k=N-1$ cases we considered in example $1.$ (given in subsection \ref{L1}) and example $2.$ (given in subsection \ref{L2}), respectively.
Furthermore, let us consider all $\binom{N}{k}$ combinations of $N$ switch positions for which there are $k$ $1$s and $N-k$ $0$s.
Thus, $M=\binom{N}{k}$.

For the time being, disregard the fact that $M$ can be astronomically large, and just assume that we can rapidly switch through all $M$ possibilities with zero switching times (this is just a thought experiement after all.)  Assuming equal times for each switch position, we have $T=\frac{N}{M} I$.

Determining $C=B^{\top} T B=\frac{N}{M}B^\top B$ is a matter of combinatorics. $B$ is a long skinny matrix $M \times N$ where $M=\binom{N}{k}$.
The entries of $B^\top B$ are the inner products of columns of $B$.

The diagonal elements of $B^\top B$ are the number of $1$s in each column. To determine this number, ask the question, given all $\binom{N}{k}$ rows with $k$ $1$s and $N-k$ $0$s, how many have a $1$ in a given position?
If the $b_j=1$ (any $j$) then in the remaining $N-1$ position, there are $k-1$ ones and $(N-1)-N-k$ $0$s. Thus, for the subvector obtained by removing bit $b_j$, there are $\binom{N-1}{k-1}$ different possibilities. Hence the diagonal elements of $B^\top B$ are $\binom{N-1}{k-1}$.

By a similar logic, the inner product of column $i$ and column $j$ is the number of rows for which there is a $b_i=1$ and $b_j=1$. This means that, in the remaining $N-2$ positions, there are $k-2$ ones and thus there are $\binom{N-2}{k-2}$ possibilities.
Thus the diagonal elements of  $B^\top B$ are $\binom{N-1}{k-1}$ and the off-diagonal elements are $\binom{N-2}{k-2}$.

Define
\begin{subequations}
\begin{IEEEeqnarray}{rCl}
M_0&=&\binom{N}{k}\\
M_1&=&\binom{N-1}{k-1}\\
M_2&=&\binom{N-2}{k-2}
\end{IEEEeqnarray}
\end{subequations}

From $T=\frac{N}{M_0} I$ we have $C=\frac{N}{M_0}\begin{bmatrix}
M_1 & M_2 & \cdots\\
M_2 & \ddots &   \\
\vdots &  & M_1 \\
\end{bmatrix}$

The diagonal elements of $C$ are
\begin{IEEEeqnarray*}{rCl}
N\frac{M_1}{M_0}&=& N \cdot \frac{\cancel{(N-k)!}(k)!}{N!} \cdot \frac{(N-1)!}{(k-1)!\cancel{(N-k)!}}\\
&=&N \cdot \frac{(k)!}{N!} \cdot \frac{(N-1)!}{(k-1)!}\\
&=&\cancel{N} \cdot \frac{k \cdot \cancel{(k-1)!}}{\cancel{N}\cdot \cancel{(N-1)!}} \cdot \frac{\cancel{(N-1)!}}{\cancel{(k-1)!}}\\
&=&k
\end{IEEEeqnarray*}

The off-diagonal elements of $C$ are
\begin{subequations}
\begin{IEEEeqnarray}{rCl}
N\frac{M_2}{M_0}&=& N \cdot \frac{\cancel{(N-k)!}(k)!}{N!} \cdot \frac{(N-2)!}{(k-2)!\cancel{(N-k)!}}\\
&=&N \cdot \frac{k (k-1)(k-2)!}{N!} \cdot \frac{(N-2)!}{(k-2)!}\\
&=&N \cdot \frac{k (k-1)(k-2)!}{N (N-1)(N-2)!} \cdot \frac{(N-2)!}{(k-2)!}\\
&=& \frac{k(k-1)}{N-1}
\end{IEEEeqnarray}
\end{subequations}

\begin{figure}[htbp] 
\centering
\includegraphics[width=\linewidth]{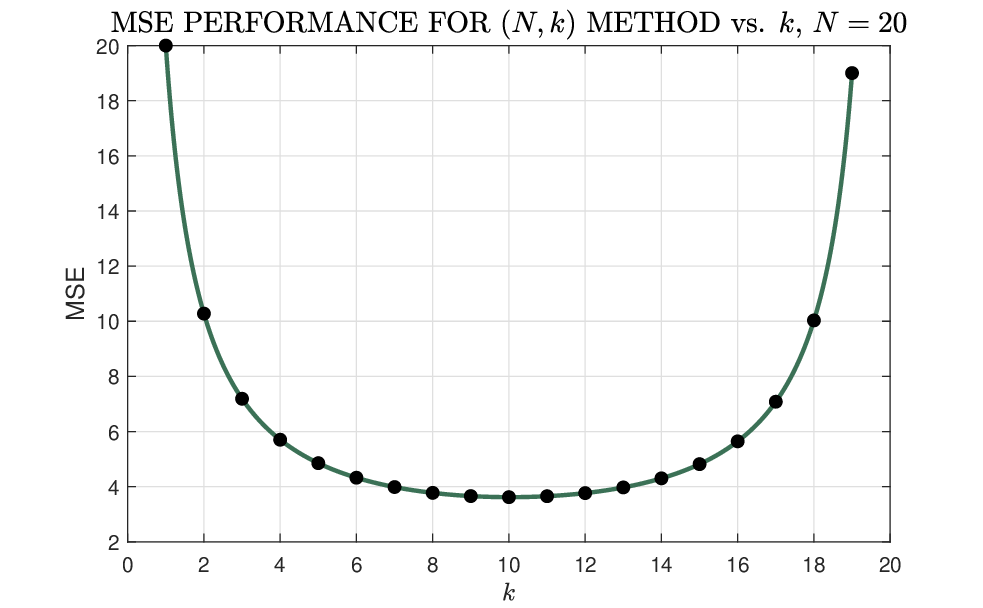}  
\caption{MSE performance for $(N,k)$ method vs. $k$, $N=20$.	}
\label{f2}
\end{figure}

Writing $$C=\begin{bmatrix}
a & b & \cdots & b\\
b & \ddots &   &\\
\vdots &  &   &\\
b &  & & a \\
\end{bmatrix}$$
\begin{IEEEeqnarray*}{rCl}
a&=&k \\
b&=&\frac{k(k-1)}{N-1}
\end{IEEEeqnarray*}
We have that $C=(a-b)I+bJ$
$$a-b=\frac{k(N-1)}{N-1}-\frac{k(k-1)}{N-1}=\frac{k(N-k)}{N-1}$$
The eigenvalues of $C$ are
\begin{IEEEeqnarray*}{rCl}
\lambda_1&=&a-b \\
&=& \frac{k(N-k)}{N-1} \quad \text{(repeated $N-1$ times)}\\
\lambda_2&=&Nb+(a-b)  \\
&=&k^2-k+k =k^2 \quad \text{(repeated $1$ time)}
\end{IEEEeqnarray*}
The eigenvalues of $C^{-1}$ are
\begin{IEEEeqnarray*}{rCl}
\lambda_1^{-1}&=& \frac{N-1}{k(N-k)} \quad \text{(repeated $N-1$ times)}\\
\lambda_2^{-1}&=&\frac{1}{k^2} \quad \text{(repeated $1$ time)}
\end{IEEEeqnarray*}
$$\operatorname{Tr} C^{-1}=\frac{(N-1)^2}{k(N-k)}+\frac{1}{k^2}$$
For the case $N=20$, a plot of $\operatorname{Tr} C^{-1}$ vs. $k$ is shown in Fig. \ref{f2}. $1 \le k\le 19$.  The function is very nearly symmetric about $\frac{N}{2}$, but not quite: the first term in $\operatorname{Tr} C^{-1}$ is symmetric about $k=\frac{N}{2}$ but the second is not.

Proposition 1.
\begin{enumerate}[label=(\alph*)]
\item $\operatorname{Tr} C^{-1}$ is a convex function of $k$.
\item For even $N$ and integer $k$ the minimum is $\frac{N}{2}$ 
\item For odd $N$ and integer $k$ the minimum will be $\frac{N+1}{2}$.
\end{enumerate}
Proof.  See Appendix A.

The even $N$, $k=\frac{N }{2}$ case is important. Let's look at $C$ in this case.
\begin{IEEEeqnarray}{rCl}
a&=&k=\frac{N}{2} \\
b&=& \frac{\frac{N}{2}(\frac{N}{2}-1)}{N-1}=\frac{\frac{1}{4}(N)(N-2)}{(N-1)} \approx \frac{N}{4}
\end{IEEEeqnarray}
The odd $N$, $k=\frac{N+1}{2}$ is also important.
\begin{IEEEeqnarray}{rCl}
a&=&k=\frac{N+1}{2} \\
b&=& \frac{\frac{N+1}{2}(\frac{N+1}{2}-1)}{N-1} \\
&=& \frac{\frac{1}{4}(N+1)(N+1-2)}{N-1}=\frac{N+1}{4}
\end{IEEEeqnarray}
\begin{IEEEeqnarray}{rCl}
C&=&\frac{N+1}{4}I+\frac{N+1}{4}J
\end{IEEEeqnarray}
Eigenvalues of $C$: 
\begin{IEEEeqnarray*}{rCl}
\lambda_1&=&\frac{N+1}{4} \quad \text{(repeated $N-1$ times)}\\
\lambda_2&=&N\frac{N+1}{4} \quad \text{(repeated $1$ time)}
\end{IEEEeqnarray*}
Eigenvalues of $C^{-1}$: 
\begin{IEEEeqnarray*}{rCl}
\lambda_1&=&\frac{4}{N+1} \quad \text{(repeated $N-1$ times)}\\
\lambda_2&=&N\frac{N+1}{4} \quad \text{(repeated $1$ time)}
\end{IEEEeqnarray*}
\begin{IEEEeqnarray}{rCl}
\operatorname{Tr} C^{-1}=\frac{N-1}{N+1}4+4 \frac{1}{N(N+1)}
\end{IEEEeqnarray}

As $N$ gets large, the optimum value of $k=\frac{N}{2}$ yields $\operatorname{Tr} C^{-1} \rightarrow 4$.  Also, $\operatorname{Tr} C^{-1}< 4$ for finite $N$.
This says that the total MSE approaches a constant, even as $N$ (and total time $T=N$) increases.  This is not the case for $k=1$, where $\operatorname{Tr} C^{-1}=N$. 
Another way to look at this result is to compare to example $4.$ (given in subsection \ref{L4}), where we observe each target individually for $t_1$ seconds, then the sum of all targets for $t_2$ seconds. If we relax the constraints on the total observation time, then $C=t_1 I+t_2 J$. This suggests that for $N$ odd, $k=\frac{N+1}{2}$, we get the same performance result as $t_1=t_2=\frac{N+1}{4}$. That is we observe each target for $\frac{N+1}{4}$ seconds, then the sum for $\frac{N+1}{4}$ seconds. The total observation time is $\frac{(N+1)^2}{4}$.

If for example $N=39$, then the total observation time is $\frac{(40)^2}{4}=400$ seconds an order of magnitude larger than $39$ seconds.

A simple rule of thumb using the bound $\operatorname{Tr} C^{-1}=4$ is that we get the same performance if we look at target individually for $\frac{N}{4}$ seconds (instead of $1$ second) so the total observation time required to get the same MSE observation with individual sensing is $\frac{N}{4}$ times longer.

\subsection{Multiple $k$ values} \label{L6}
The result above says that if our strategy is to close $k$ switches in each observation interval and we consider all $M=\binom{N}{k}$ combinations of switch positions, then the optimal value of $k$ is $\frac{N}{2}$ for $N$ even and $\frac{N+1}{2}$ for $N$ odd.

This result can be extended by considering multiple $k$ values, allocating equal time for all $\binom{N}{k}$ configurations of $k$ switches closed, but allowing different times for different values of $k$.  Suppose we take our total time $N$ and divide it into $N$ time segments $\alpha_k N$, with $\sum_{k=1}^N \alpha_k = 1$ and in the $k^{th}$ time segment we observe all possibilities of $k$ switches closed.  The total number of observations in this scenario could be as large as $2^N -1$, if all the $\alpha_k$ are nonzero.

Let $$B=\begin{bmatrix}
B_1\\
B_2\\
\vdots\\
B_N\\
\end{bmatrix}$$
where $B_k$ is the $\binom{N}{k} \times N$ matrix whose rows are binary vectors of length $N$ with $k$ 1s.  Further define \begin{IEEEeqnarray}{rCl}
C_k=B_k^{\top}T_kB_k
\end{IEEEeqnarray}
where $T_k$ is the diagonal matrix of times allocated to the observations with $k$ switches closed.   For now we take $T_k=\alpha_kNI$.  Then 
\begin{IEEEeqnarray}{rCl}
C = \sum_{k=1}^nC_k = N\sum_{k=1}^N \alpha_kB_k^{\top}B_k .
\end{IEEEeqnarray}
That is, the matrix $C$ is a convex combination of matrices that would result from observations with a single $k$, described in the subsection above.  We have already seen that 
\begin{IEEEeqnarray}{rCl}C_k = a_kI + b_kJ,
\end{IEEEeqnarray} 
and hence 
\begin{IEEEeqnarray}{rCl}
C = (\sum \alpha_k a_k)I + (\sum \alpha_k b_k) J .
\end{IEEEeqnarray}

All the $C_k$ have the same set of eigenvectors, which are shared by $C$.  The eigenvalues are $$\lambda_1 = \sum \alpha_k \frac{k(N-k)}{N-1}    \text{(repeated $N-1$ times)}$$ and $$\lambda_2 = \sum \alpha_k k^2    \text{(repeated $1$ time)} .$$

The main result of this subsection is the following.

Proposition 2.  For matrices $C$ with the structure described here, the MSE is minimized when only a single value of $k$ is used, that value being $k = \frac{N}{2}$, $N$ even and $N \ge 4$, and $k= \frac{(N+1)}{2}$, all odd $N \ge 3$.  The case $N=2$ is a notable exception.

Proof.  See Appendix B.  

\begin{figure}[t] 
\centering
\includegraphics[width=\linewidth]{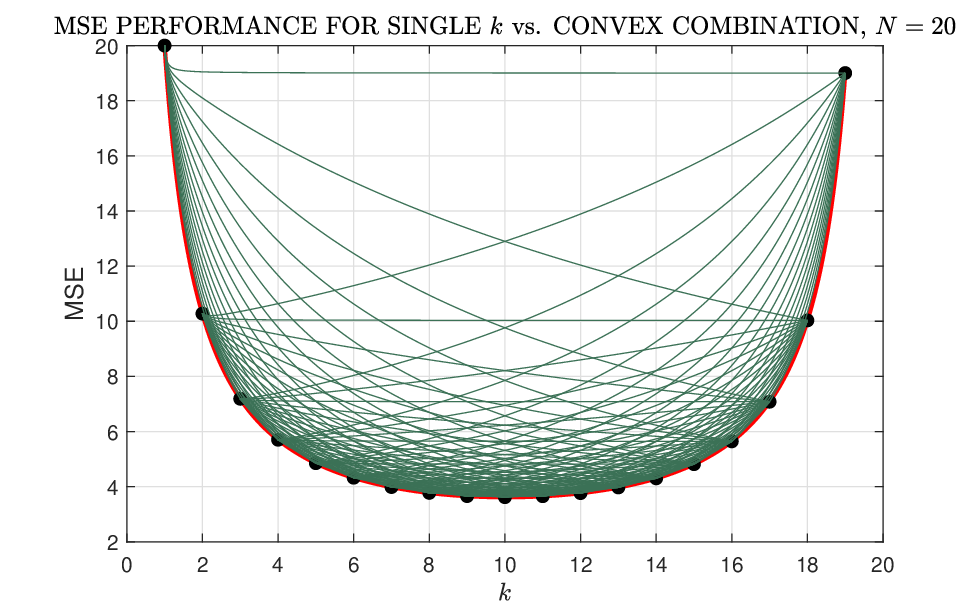}  
\caption{MSE performance for single $k$ vs. convex combination, $N=20$. 	}
\label{f3}
\end{figure}

The proof is somewhat complicated, but as an illustration of the basic idea, in Fig. \ref{f3} we show all the MSE values that result for $N=20$ by taking two $k$ values at a time and computing the MSE for all convex combinations.  The curves shown are not straight lines; they are convex functions that lie above the curve first shown in Fig. \ref{f2}.

\subsection{Extension to Arbitrary $C$} \label{L7}

We can now complete the main result of this section, which is that the solutions found in example $1.$ (in subsection \ref{L1}) above, with a single value of $k$ switches closed, and equal time allocated to all $\binom{N}{k}$ switch configurations, minimizes the MSE over all possible values of times in the diagonal matrix $T$, again with the exception of $N=2$.  

Proposition 3. Let $C = B^{\top}TB = \sum_{k=1}^N \alpha_k B_k^{\top}T_kB_k$ where all terms are defined above.  Then $\operatorname{Tr}C^{-1}$ is minimized over all choices of $T$ for $T_k = \frac{N}{\binom{N}{k}} I$ for $k = \frac {N}{2}$, $N$ even and $N \ge 4$, and $k = \frac{(N+1)}{2}$, $N$ odd and $N \ge 3$, and all other $T_i = 0$.  

Proof.  See Appendix C. The basic idea is that the largest eigenvalue of any $C$ with this construction is greater than or equal to the largest eigenvalue of a matrix $C$ with the equal-time allocation constraint of example $2.$ (in subsection \ref{L2}). By reducing this eigenvalue and making all the smaller eigenvalues equal, one can create a new matrix $C'$ whose eigenvalues coincide with those of a matrix from example $2.$ (subsection \ref{L2}), and for which $\operatorname{Tr}C'^{-1}$ is reduced.  In Proposition $2$, we have already found that the minimizer over all such matrices is the one given in the statement of Proposition $3$, hence this value of $\operatorname{Tr}C^{-1}$ found previously is the minimum over all possible values of $C$, with the $N=2$ case being an exception.

\section{Hadamard matrix construction} \label{S5}

The obvious flaw in the data collection strategy described above is that for even moderate values of $N$, $\binom{N}{N/2}$ grows factorially and any kind of practical data collection, storage and processing becomes impossible.
For example, if $N=40$ then $\binom{40}{20}=1.3785 \times 10^{11}$. If $40$ seconds are available for data collection, each time segment would be $\frac{40}{\binom{40}{20}} \approx 0.29 $ns, meaning we would be resetting the switches at the integrator every $\frac{1}{3}$ns and would have to store and process $138$Gsamples of data. Clearly it would be difficult to find a practical application of this result (even if it is only a thought experiment) and the situation only gets worse as $N$ grows.

Fortunately, there is a way out.  We have found a square matrix $B$ for certain values of $N$ that achieves exactly the same MSE performance as the optimal $B$ described above with $\binom{N}{k}$ rows.

\subsection{Basic Hadamard construction}\label{L8}

The construction involves Hadamard matrices \footnote{The Hadamard matrix conjecture, one of the open problems in discrete mathematics, states a Hadamard matrix exists for any order that is a multiple of $4$. This still remains unproven but Hadamard matrices for all but a few orders less than $2000$ are known. The smallest order for which no Hadamard matrix has been found is $668$.}. We assume that $N$ is odd and $N+1$ is a multiple of $4$ and that a Hadamard matrix of order $N+1$ exists.

A Hadamard matrix $H$ is a matrix whose entries are $+1$ and $-1$ and is orthogonal:
\begin{IEEEeqnarray}{rCl}
H^\top H=H H^\top =N I.
\end{IEEEeqnarray}
For any Hadamard matrix $H$, $DH$ and $HD$ are also Hadamard matrices for $D$ being a diagonal matrix with $+1$ and $-1$ elements on the diagonal (representing inversion of rows /columns). Thus, we can assume without loss of generality that the first row and column of $H$ is all-ones. Such as $H$ is called normal Hadamard matrix. 

Let $H$ be a normal Hadamard matrix of size $N+1$, and let $\tilde{H}$ be the $N \times N$ matrix obtained by deleting the first row and column. Note that each row and column of $\tilde{H}$ as $\frac{N+1}{2} (-1)$s and $\frac{N-1}{2} (+1)$s.

The norm of each row/column is $N$ and the inner product of pairs of rows/columns is $-1$. Hence
\begin{IEEEeqnarray}{rCl}
\tilde{H}^\top \tilde{H}=\tilde{H}\tilde{H}^\top=\begin{bmatrix}
N& -1 & \cdots & -1\\
-1 & \ddots &   \\
\vdots & &  &  \\
-1 & & & N \\
\end{bmatrix}=(N+1)I-J
\end{IEEEeqnarray}
The matrix $B$ that meets our needs is 
\begin{IEEEeqnarray}{rCl}
B=\frac{1}{2}(J-\tilde{H}).
\end{IEEEeqnarray}
That is, to obtain $B$, take $\tilde{H}$ and convert every $-1$ to $+1$ and every $1$ to $0$.  We have that
\begin{subequations}
\begin{IEEEeqnarray}{rCl}
B^\top B &=& \frac{1}{2}(JJ-J\tilde{H}-H^\top J+ \tilde{H}^\top H)\\
&=& \frac{1}{4}(NJ+J+J+(N+1)I-J)\\
&=& \frac{1}{4}\big((N+1)J+(N+1)I \big)\\
&=& \frac{N+1}{4}I+\frac{N+1}{4}J
\end{IEEEeqnarray}
\end{subequations}
This is exactly the same $C$ matrix obtained in section \ref{S4} above, with a $B$ matrix with $\binom{N}{\frac{N+1}{2}}$ rows.

In sum, a square matrix $B$ (and hence $T=I$) that achieves the same MSE performance as the best obtained previously, is found by the following algorithm:
\begin{enumerate}[label=(\alph*)]
\item compute Hadamard matrix of order $N+1$
\item delete $1$st row and column
\item covert $-1$s to $+1$s and $+1$s to $0$s.
\end{enumerate}
\begin{IEEEeqnarray}{rCl}
B^\top B =\frac{N+1}{4}I+\frac{N+1}{4}J
\end{IEEEeqnarray}
and the MSE is identical to that obtained previously.

This matrix $B$ (and its complement $J-B$) appears often in a field called "combinatorial design theory," which apparently has many applications, for example, as error-correcting code \cite{design}.  We are motivated to find other interesting signal processing applications of design theory, not known to the authors prior to this work.

\subsection{Unequal observations times} \label{L9}

Now that we have a practical solution to the observation problem that involves only $N$ observations, we can again ask the question, is there any advantage to be gained by considering unequal observation times, that is by using a diagonal matrix $T=\operatorname{diag}(t_1 \cdots t_N)$ where $\sum_{i=1}^{N}b_i=N$ but not all $t_1=1$?

The answer is no, and the proof technique closely mirrors what was used in subsection \ref{L3} above and in the proof of Proposition $3$.

Consider the quadratic form $u^{\top} C u$ where $u$ is any unit-norm vector.  It is well-known that the largest eigenvalue of $C$ maximizes this quadratic form over all choices of $u$.  Take $u = \frac{1}{\sqrt{N}} 1$, where $1$ denotes the all-ones vector. It is easily seen that
\begin{IEEEeqnarray}{rCl}
B1 = (\frac{N+1}{2})1
\end{IEEEeqnarray}
and hence 
\begin{IEEEeqnarray}{rCl}
u^{\top}B^{\top}TBu = \frac{1}{N} (\frac{N+1}{2})^2 \operatorname{Tr} T = (\frac{N+1}{2})^2 .
\end{IEEEeqnarray}

We have found a unit-norm vector $u$ $\operatorname{s.t.}$ $u^\top Cu =$ largest eigenvalue of $B^\top B$, which means that the largest eigenvalue $(C) >$ largest eigenvalue of $(B^\top B)$.
This implies there is a transformation of the eigenvalues of $C$ that takes them back to the eigenvalues of $(B^\top B)$.

\setlist[enumerate]{label={\arabic*.}} 
\begin{enumerate}
\item reduce the largest eigenvalue of $\tilde{C}$ to $\big(\frac{N+1}{2}\big)^2$
\item take the "excess" and redistribute to the smaller eigenvalues
\item replace the smaller eigenvalues by their average.
\end{enumerate}

The operators $(1-3)$ above represent a doubly stochastic operation on the eigenvalues of $C$, resulting in the eigenvalues of $B^\top B$.
Since $\operatorname{Tr} C^{-1}=\sum_{i=1}^{N}\frac{1}{\lambda_i}$ and $f(t)=\sum_{i=1}^{N}\frac{1}{\lambda_i}$ is a Schur convex function (symmetric too), it follows that $\operatorname{Tr} C^{-1} > \operatorname{Tr} {(B^\top B)}^{-1}$.
Hence having equal observation times $(T=I)$ is optimal:
\begin{IEEEeqnarray}{rCl}
\operatorname{Tr} {(B^\top T B)}^{-1} > \operatorname{Tr} {(B^\top B)}^{-1}
\end{IEEEeqnarray}
for $T \neq I$, $\operatorname{Tr} T=N$.
\subsection{Extension to all $N$}\label{L10}

For $N+1$ not a multiple of $4$, there is still strategy based on the Hadamard design.

For any $N$, take $(N+j)$ to be the next largest multiple of $4$. 
Compute the $(N+j) \times (N+j)$ Hadamard matrix $H$. Let $\tilde{H}=(N-1+j) \times (N-1+j)$ matrix with $1$st row/column removed. $\tilde{B}=\frac{1}{2 }(J-\tilde{H})$ as before.

\begin{figure}[t] 
\centering
\includegraphics[width=\linewidth]{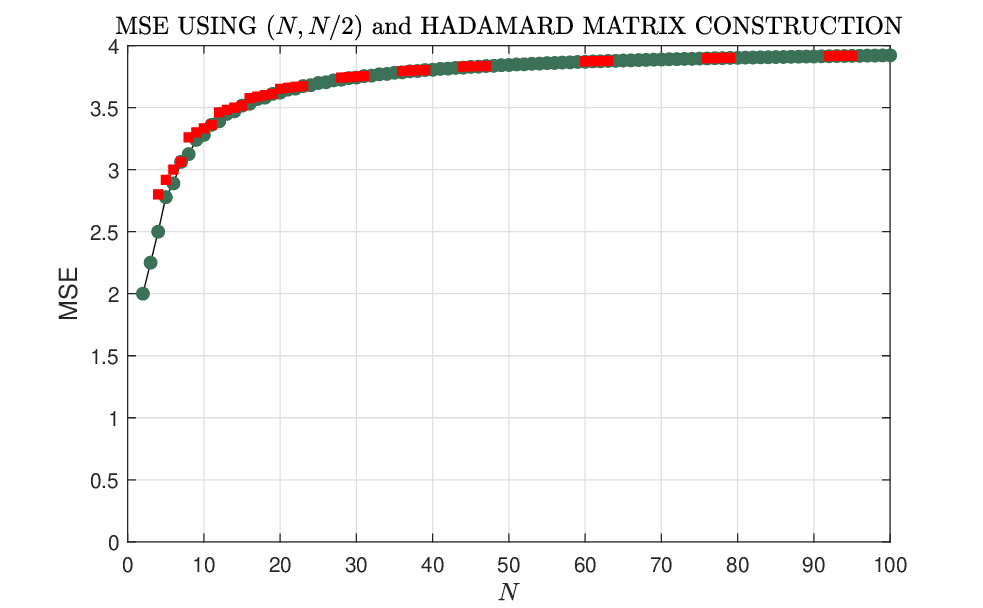}  
\caption{MSE using $(N,\frac{N}{2})$ and Hadamard matrix construction. }
\label{f4}
\end{figure}

Let $B=\tilde{B}$ with the rightmost $J-1$ columns removed. Hence $B$ is $(N+j-1) \times N$ (it is rectangular).
$B^\top B$ has the same $aI+bJ$ structure as before:
\begin{IEEEeqnarray}{rCl}
C=\frac{N}{N-j+1}B^\top B
\end{IEEEeqnarray}
$\operatorname{Tr} C^{-1}$ is very close to what was computed previously, as illustrated in Fig. \ref{f4} for even values of $N$ and $k = \frac{N}{2}$.  The green circles show the optimum MSE using the $M=\binom{N}{k}$.  Superimposed on this plot using red squares in groups of four is the MSE for the solution proposed above based on the Hadamard design, with the last square in each group corresponding to $N+1$ being a multiple of four and hence coinciding with a green circle exactly.  Note how, for $N$ greater than $20$ or $30$, all the red squares are indistinguishable from the green circles, at least given the resolution of this graphic.

\subsection{Relationship to Walsh Codes and CDMA}\label{L11}
Our approach to combining multiple signals through a network of switches, which in effect is equivalent to multiplying them by $0$s and $1$s, closely resembles other methods for combining signals in a multiple-access communication system.   If we consider the columns of the $B$ matrix, and the times $T$ spent in each switch position, one has a set of signals that take on the values $0$ and $1$, and each signal multiplies one of the $\mu$ values before they are added together.  

If all the $T$ values are equal, and the $B$ values were $+1$ and $-1$ (not $0$ and $1$), then the optimal $B$ would be a Hadamard matrix, assuming one exists for the given value of $M$.   It is not difficult to show that such a matrix maximizes both SNR and signal separation, under the constraint that the $B$ values are less than or equal to $1$ in absolute value.  Such a Hadamard matrix is known as a Walsh code, after the mathematician J. L. Walsh who introduced a set of closely related continuous-time functions in $1923$ \cite{walsh1923closed}.   The Walsh code is used extensively in Code Division Multiple Access (CDMA) digital communication systems.  In such communication systems, the digital data are mapped into the $\mu$ values which would take values in a finite alphabet, although methods which minimize MSE in estimating $\mu$ on a continuum would play directly into the symbol detection problem for finite alphabets.   

What we have done is identify signals that take on the values $0$ and $1$, such as on-off keying (OOK) signals, with similar optimality properties. Obviously multiple OOK signals cannot be orthogonal but we can still choose them to minimize the MSE of the signal estimates.  It is interesting that the optimal solution involves Hadamard matrices, just as in the case of the Walsh codes.  We also note that there is a distinction between Walsh codes and Hadamard codes, which are error-correcting codes based on a conversion of a Hadamard matrix to a $0-1$ matrix.   We have not explored any aspects of error correction in the current problem although such a connection probably exists.  Both Walsh codes for CDMA and Hadamard codes for error correction are discussed in Chapter $3$ of \cite{horadam2007hadamard}. 

\section{Summary and Conclusion}
We have presented an abstraction for multisensor communication which we
all the Gaussian Multiplex Channel.  In this model there are $N$ sensor
outputs which can be added together in any combination through a network
of switches, and the combinations can be changed arbitrarily throughout
the observation interval.  The sensor output sums are observed in
additive Gaussian noise.  Using a MSE cost function, we have solved the
optimization problem which yields one possibility for the switch
positions and observation times which minimizes the MSE given a
constraint on the total observation.  While this first solution
identifies the global minimum of the MSE over all possible observation
schemes, it has extraordinarily high complexity - $ \binom{N}{N/2}$ different switch positions - even for moderate values of $N$.

While our first solution does identify a high-complexity solution which
achieves the global minimum of the MSE, it does not preclude the
possibility of other lower-complexity solutions which achieve the same
MSE.  Such a solution does exist, based on Hadamard designs, which
requires a number of combinations equal to the number of sensors, for
the case when $N+1$ is a multiple of $4$.  For other values of $N$, there is a
closely related solution which is still very nearly optimal.  This
significantly increases the chances that there may be some useful
application of our results.

The motivating technology for our abstraction is the proliferation of
large numbers of sensors in an environment with limited communication
capacity.  Some adaptation and modification of the original abstract model may be required for specific engineering applications.
\bibliographystyle{IEEEtran}
\bibliography{main}
\appendices
\section{Proof of proposition 1}
\begin{enumerate}[label=(\alph*)]
\item 
\begin{subequations}
\begin{IEEEeqnarray}{rCl}
f(k)&=& \frac{(N-1)^2}{k(N-k)} +\frac{1}{k^2}\\
&=& \frac{(N-1)^2/N}{k} +\frac{(N-1)^2/N}{N-k}+\frac{1}{k^2}
\end{IEEEeqnarray}
\end{subequations}
$f(k)$ is the sum of three convex functions, hence it is convex.

For (b) and (c), by the convexity of $k$, it suffices to show that $f(k) < f(k-1)$ and $f(k) < f(k+1)$ at the minimizing integer $k_s$ i.e. $f(k)$ is less than its neighbors on the left and right.

\item $N$ even.
$N=2$ is a trivial special case, since $k=1$ the only integer value of $k$ for which $f(k)$ is finite.

Otherwise, for $N >2:$

\begin{IEEEeqnarray}{rCl}
f\Big(\frac{N}{2}\Big)&=& \frac{(N-1)^2}{(\frac{N}{2})^2} +\frac{1}{\big(\frac{N}{2}\big)^2}\\
f\Big(\frac{N}{2}-1\Big)&=& \frac{\big(N-1\big)^2}{\big(\frac{N}{2}-1\big) \big(\frac{N}{2}+1\big)} +\frac{1}{\big(\frac{N}{2}-1\big)^2}\\
f\Big(\frac{N}{2}+1\Big)&=& \frac{\big(N-1\big)^2}{\big(\frac{N}{2}-1\big) \big(\frac{N}{2}+1\big)} +\frac{1}{\big(\frac{N}{2}+1\big)^2}
\end{IEEEeqnarray}
$f\Big(\frac{N}{2}-1\Big)>f\Big(\frac{N}{2}\Big)$, as the denominators in both terms of $f\Big(\frac{N}{2}-1 \Big)$ are greater than their counterparts in $f\Big( \frac{N}{2}  \Big) .$

$f\Big(\frac{N}{2}+1\Big)$ requires more calculation
\begin{IEEEeqnarray*}{rCl}
f\Big(\frac{N}{2}+1\Big)-f\Big(\frac{N}{2} \Big) = \frac{\big(N-1\big)^2}{\big(\frac{N}{2}-1\big) \big(\frac{N}{2}+1\big)} +\frac{1}{\big(\frac{N}{2}+1\big)^2}-\frac{\big(N-1\big)^2}{\big(\frac{N}{2}\big)^2} -\frac{1}{\big(\frac{N}{2}\big)^2} \IEEEyesnumber
\end{IEEEeqnarray*}
Multiply by $\frac{1}{4},$ and reorganizing  this becomes
\begin{IEEEeqnarray}{rCl}
&&(N-1)^2 \Big[ \frac{N^2-(N^2-4)}{(N-2)(N+2)N^2}  \Big] - \Big[ \frac{N^2+4N+4-N^2)}{N^2(N+2)^2}  \Big] \IEEEnonumber\\
=&&(N-1)^2 \Big[ \frac{4}{(N-2)(N+2)N^2}  \Big] - \Big[ \frac{4(N+1)}{N^2(N+2)^2}  \Big]
\end{IEEEeqnarray}
Now multiply by $\frac{N^2(N+2)(N-2)}{4(N-1)}$
(which is positive for $N>2$):
\begin{IEEEeqnarray}{rCl}
&&(N-1)-\frac{(N+1)(N-2)}{(N+2)(N-1)} \IEEEnonumber\\
=&&(N-1)-\frac{(N^2-N-2)}{(N^2+N-2)}
\end{IEEEeqnarray}
The term on the right is positive and less than $1$ for all $N \ge 4$, hence the expression is positive and $f(\frac{N}{2}+1) > f(\frac{N}{2})$.

\item $N$ odd 
\begin{IEEEeqnarray}{rCl}
f\Big(\frac{N+1}{2}\Big)&=& \frac{(N-1)^2}{(\frac{N+1}{2})(\frac{N-1}{2})} + \frac{1}{\big(\frac{N+1}{2}\big)^2} \\
f\Big(\frac{N-1}{2}\Big)&=& \frac{(N-1)^2}{(\frac{N+1}{2})(\frac{N-1}{2})} + \frac{1}{\big(\frac{N-1}{2}\big)^2} \\
f\Big(\frac{N+3}{2}\Big)&=& \frac{(N-1)^2}{(\frac{N+3}{2})(\frac{N-3}{2})} + \frac{1}{\big(\frac{N+3}{2}\big)^2}
\end{IEEEeqnarray}
We see that $f(\frac{N-1}{2}) >f(\frac{N+1}{2})$, since the two first terms are equal, and the second term in $f(\frac{N-1}{2})$ is greater than its counterpart in $f(\frac{N+1}{2})$.

$f(\frac{N+3}{2})$ requires more calculation.

$N=3$ is a special case since $f(3)$ is not finite and hence $f(2)<f(3)$ immediately. Otherwise:
\begin{IEEEeqnarray}{rCl}
&&f\Big(\frac{N+3}{2}\Big)-f\Big(\frac{N+1}{2}\Big)=\IEEEnonumber\\
&&\frac{(N-1)^2}{(\frac{N+3}{2})(\frac{N-3}{2})}+\frac{1}{(\frac{N+3}{2})^2}-\frac{(N-1)^2}{(\frac{N+1}{2})(\frac{N-1}{2})}-\frac{1}{(\frac{N+1}{2})^2}
\end{IEEEeqnarray}
Multiplying by $\frac{1}{4}$ and reorganizing yields
\begin{subequations}
\begin{IEEEeqnarray}{lll}
&&(N-1)^2 \Big[ \frac{1}{(N+3)(N-3)}-\frac{1}{(N+1)(N-1)}  \Big]- \Big[ \frac{1}{(N+1)^2}-\frac{1}{(N+3)^2}\Big] \IEEEnonumber \\
=&& (N-1)^2 \Big[ \frac{(N^2-1)-(N^2-9)}{(N+3)(N-3)(N+1)(N-1)}     \Big] -\Big[\frac{(N+3)^2-(N+1)^2}{(N+1)^2(N+3)^2}\Big] \\
=&& (N-1)^2 \Big[ \frac{8}{(N+3)(N-3)(N+1)(N-1)}     \Big] -\Big[\frac{4(N+2)}{(N+1)^2(N+3)^2}\Big]
\end{IEEEeqnarray}
\end{subequations}
Multiplying by $\frac{(N+3)(N-3)(N+1)}{8}$ which is positive for $N>3$ yields
\begin{IEEEeqnarray}{rCl}
&&(N-1)- \Big( \frac{(N+2)(N-3)}{(N+1)(N+3)}       \Big) \frac{1}{2} \IEEEnonumber \\
=&&(N-1)-\Big( \frac{N^2-N-6}{N^2+4N+3}       \Big) \frac{1}{2}
\end{IEEEeqnarray}
The term on the right is positive and less than $1$ for $N \ge 5$, hence the overall expression is positive and $f(\frac{N+3}{2}) > f(\frac{N+1}{2})$.
\end{enumerate}

\section{Proof of proposition 2}
Define $k_*=\frac{N}{2}$ ($N$ even) or $\frac{N+1}{2}$ ($N$ odd).
Define $C_*$ as the corresponding matrix $C$

Let $0 \le \beta \le 1$, further let $\alpha_k$ be defined for all $k \ne k_*$ $\operatorname{s.t.} \alpha_k \ge 0$ and $\sum_{k \ne k_*} \alpha_k=1$.
Let $(1-\beta)$ be the weight on $C_*$, and $\beta \alpha_k$ the weights on the other $C_k$. Then
\begin{IEEEeqnarray}{rCl}
C=(1-\beta)C_* + \beta \sum_{k\ne k_*}\alpha_k C_k
\end{IEEEeqnarray}
The eigenvalues of $C$ are
\begin{IEEEeqnarray}{rCl}
\lambda_1&=&(1-\beta) \frac{k+(N-k_*)}{N-1}+\beta \sum_{k \ne k_*} \alpha_k \frac{k(N-k)}{N-1} \quad \text{(repeated $N-1$ times)} \\
\lambda_2&=&(1-\beta) k_*^2+\beta \sum_{k \ne k_*} \alpha_k k^2 \quad \text{(repeated one time)}
\end{IEEEeqnarray}
Then
\begin{IEEEeqnarray}{rCl}
\operatorname{Tr}C^{-1}=\frac{N-1}{\lambda_1}+\frac{1}{\lambda_2}
\end{IEEEeqnarray}
Our general approach will be to show that either
\setlist[enumerate]{label={\arabic*.}} 
\begin{enumerate}
\item $\operatorname{Tr}C^{-1}>\operatorname{Tr}C_*^{-1}$, immediately or
\item $\frac{\partial}{\partial \beta}\operatorname{Tr}C^{-1} > 0$, all $0 \le \beta \le 1$
\end{enumerate}

$N$ even:
\begin{IEEEeqnarray}{rCl}
\operatorname{Tr}C^{-1}=\frac{(N-1)^2}{(1-\beta)(\frac{N}{2})^2+\beta \sum a_k k (N-k)}+\frac{1}{(1-\beta)(\frac{N}{2})^2+\beta \sum \alpha_k k^2}
\end{IEEEeqnarray}

Note that $\sum a_k k (N-k) < (\frac{N}{2})^2$, by Convexity.
If $\sum \alpha_k k^2 < (\frac{N}{2})^2$, then immediately
\begin{IEEEeqnarray}{rCl}
\operatorname{Tr} C^{-1} > \operatorname{Tr} C^{-1}_*
\end{IEEEeqnarray}

Since both denominators in $\operatorname{Tr} C^{-1}$ are less than their counterparts in $\operatorname{Tr} C^{-1}_*$ Otherwise the first denominator decreases w.r.t $\beta$ but the second denominator increases, and we must consider $\frac{\partial}{\partial \beta}\operatorname{Tr} C^{-1}$.
Define
\begin{IEEEeqnarray}{rCl}
\gamma&=&\sum \alpha_k k (N-k) \\
\delta&=& \sum a_k k_z
\end{IEEEeqnarray}
Also define $D_1$ and $D_2$ as the denominators of the two terms in $\operatorname{Tr}C^{-1}$, respectively.
We have
\begin{IEEEeqnarray}{rCl}
\frac{\partial}{\partial \beta} \operatorname{Tr}C^{-1} = \frac{(N-1)((\frac{N}{2})^2- \gamma)}{D_1^2}+\frac{(\frac{N}{2})^2-\delta}{D_2^2}
\end{IEEEeqnarray}
Since $\gamma <(\frac{N}{2})^2$ and $\delta>(\frac{N}{2})^2$, we have $D_1^2<D_2^2$ and hence
$$\frac{\partial}{\partial \beta} \operatorname{Tr}C^{-1} > \frac{(N-1)((\frac{N}{2})^2- \gamma)+(\frac{N}{2})^2-\delta}{D_2^2}$$
It suffices to show the numerator is positive. This is
\begin{IEEEeqnarray}{rCl}
(N-1)^2 \Big(\big(\frac{N}{2}\big)^2-\sum\alpha_k k (N-k)\Big)+\Big(\frac{N}{2}\Big)^2-\sum\alpha_k k^2
\end{IEEEeqnarray}
Using $(\frac{N}{2})^2=\sum \alpha_k (\frac{N}{2})^2$ this can be written as
\begin{IEEEeqnarray}{rCl}
\sum a_k \Big((N-1)^2\Big(\big(\frac{N}{2}\big)^2-Nk+k^2\Big)+\big(\frac{N}{2}\big)^2-k\Big)
\end{IEEEeqnarray}
To show this is positive for any combination of $a_k$, it must be that every multiplier of $a_k$ is positive. If $k<\frac{N}{2}$ the term is positive. otherwise for $k>\frac{N}{2}$ we have
$$(N-1)^2\Big(k-\frac{N}{2}\Big)^2 +\Big(\frac{N}{2}-k\Big)\Big(\frac{N}{2}+k\Big)$$
Dividing by $\big(k-\frac{N}{2} \big)^2$ we get
\begin{IEEEeqnarray}{rCl}
(N-1)^2 -\Big(k+\frac{N}{2}\Big)\Big/\Big(k-\frac{N}{2}\Big)
\end{IEEEeqnarray}
The second term is largest for integer $k>\frac{N}{2}$ when $k=\frac{N}{2}+1$, and this becomes $(N-1)^2-\frac{N+1}{1}=N(N-3)$.
This is $>0$ for all even $N \ge 4$. Note that $N=2$ is a special case for which the general result does not hold.

So, for all even $N \ge 4$, $\frac{\partial}{\partial \beta}\operatorname{Tr}C^{-1}>0$, meaning that the $\beta=0$ solution, i.e. $C_*$, minimizes $\operatorname{Tr}C^{-1}$.

Odd $N$

The approach is similar
\begin{IEEEeqnarray}{rCl}
\operatorname{Tr}C^{-1}&=& \frac{(N-1)^2}{(1-\beta)(\frac{N+1}{2})(\frac{N-1}{2})+\beta \sum \alpha_kk(N-k)}+\frac{1}{(1-\beta)(\frac{N+1}{2})^2+\beta \sum \alpha_k k^2} \\
&=& \frac{(N-1)^2}{(1-\beta)(\frac{N+1}{2})(\frac{N-1}{2})+\beta \gamma}+\frac{1}{(1-\beta)(\frac{N+1}{2})^2+\beta \delta}
\end{IEEEeqnarray}
$\gamma < \big( \frac{N+1}{2}\big)\big( \frac{N-1}{2}\big)$ by convexity. If $\delta < \big( \frac{N+1}{2}  \big)^2$, then $\operatorname{Tr}C^{-1}>\operatorname{Tr}C^{-1}_*$ since both denominators decrease when $\beta \ne 0$.

Otherwise consider $\frac{\partial}{\partial \beta} \operatorname{Tr}C^{-1}$:
\begin{IEEEeqnarray}{rCl}
\frac{\partial}{\partial \beta} \operatorname{Tr}C^{-1} = \frac{(N-1)^2((\frac{N+1}{2})(\frac{N-1}{2})- \gamma)}{D_1^2}+\frac{(\frac{N+1}{2})^2-\delta}{D_2^2}
\end{IEEEeqnarray}
Again $\delta >\big( \frac{N+1}{2}  \big)^2$ we have $D_1^2 <D_2^2$
\begin{IEEEeqnarray}{rCl}
\frac{\partial}{\partial \beta} \operatorname{Tr}C^{-1} > \frac{(N-1)^2 \big((\frac{N+1}{2})(\frac{N-1}{2})- \gamma\big)+(\frac{N+1}{2})^2-\delta}{D_2^2}
\end{IEEEeqnarray}
The numerator is
\begin{IEEEeqnarray}{rCl}
\sum \alpha_k \Big[ (N-1)^2\Big(\Big(\frac{N+1}{2}\Big)\Big(\frac{N-1}{2}\Big)-kN+k^2+ \Big(\frac{N+1}{2}\Big)^2-k^2 \Big]
\end{IEEEeqnarray}

The multiplier for $\alpha_k$ is
\begin{IEEEeqnarray}{rCl}
(N-1)^2 \Big(k-\frac{N+1}{2}\Big)\Big(k-\frac{N-1}{2}\Big)+\Big(\frac{N+1}{2}-k\Big)\Big(\frac{N+1}{2}+k\Big)
\end{IEEEeqnarray}
Divide by $\Big(k-\frac{N+1}{2}\Big)\Big(k-\frac{N-1}{2}\Big)$ (which is positive)
\begin{IEEEeqnarray}{rCl}
(N-1)^2-\frac{k+\frac{N+1}{2}}{k-\frac{N-1}{2}}
\end{IEEEeqnarray}
The second term is maximized (most negative) for $k>\frac{N+1}{2}$ at $k=\frac{N+1}{2}+1$
\begin{IEEEeqnarray}{rCl}
(N-1)^2-\frac{N+2}{2} \implies 2(N-1)^2-N+2=2N^2-5N+4
\end{IEEEeqnarray}
This is $\ge0$ for all $N \ge 3$
\begin{IEEEeqnarray}{rCl}
\implies \frac{\partial}{\partial \beta} \operatorname{Tr}C^{-1} >0  \implies \operatorname{Tr}C^{-1} > \operatorname{Tr}C^{-1}_*
\end{IEEEeqnarray}

\section{Proof of proposition 3}
Write $C$ as
\begin{IEEEeqnarray}{rCl}
C=\sum_{k=1}^{N} \alpha_{k} C_{k}
\end{IEEEeqnarray}
where $\alpha_k \ge 0,$ $\sum_{k=1}^{N} \alpha =1$
\begin{IEEEeqnarray}{rCl}
C_k=B^\top_k T_k B_k
\end{IEEEeqnarray}
\begin{IEEEeqnarray}{rCl}
\operatorname{Tr} T_k=N
\end{IEEEeqnarray}
We note that 
\begin{IEEEeqnarray}{rCl}
\operatorname{Tr}C_k&=&\operatorname{Tr} T_k(B_k B^\top) \\
&=&Nk
\end{IEEEeqnarray}
The proof relies on the well-known fact the quadratic form $u^\top C u$ is maximized over all unit-norm $u$ at the maximum eigenvalue of $C$.

Let $u=\frac{1}{\sqrt{N}} 1_N$, where $1_N$ is the $N \times 1$ all-ones vector
\begin{IEEEeqnarray}{rCl}
u^\top C u&=&\sum_{k=1}^{N} \alpha_k u^\top B_k T_k B_k u \\  
&=& \frac{1}{N} \sum_{k=1}^{N} \alpha_k 1^\top_N B_k T_k B_k 1_N\\
B_k 1_N&=&k 1_{\binom{N}{k}}
\end{IEEEeqnarray}
where now the vector $1_{\binom{N}{k}}$ on the r.h.s is $\binom{N}{k} \times 1$.
\begin{IEEEeqnarray}{rCl}
1^\top _N B^\top_k T_k B_k 1_N&=& k^2 1^\top_{\binom{N}{k}}T_k 1_{\binom{N}{k}}^\top   \\  
k^2 \operatorname{Tr} T_k&=&k^2 N
\end{IEEEeqnarray}
Hence $u^\top C u=\sum_{k=1}^{N} \alpha_k k^2$.

This says that the largest eigenvalue of $C$ is greater than or equal to $\sum_{k=1}^{N} \alpha_k k^2.$

We can construct a new matrix $C'$ by performing the following operations:
\begin{enumerate}[label=(\alph*)]
\item reduce the maximum eigenvalue to $\sum_{k=1}^{N} \alpha_k k^2$
\item distribute the "excess" to the $N-1$ smaller eigenvalues
\item replace the $N-1$ smaller eigenvalues with their average. Note that this is guaranteed to be smaller than $\lambda_{\text{max}}$ by $\operatorname{Tr}C=\operatorname{Tr} C'=N \sum_{k=1}^N \alpha_k k$.
\end{enumerate}

This sequence of operations represents a doubly stochastic transformation of the eigenvalues of $C.$ Since $\operatorname{Tr}C^{-1}$ is a Schur-convex of the eigenvalues, we have
\begin{IEEEeqnarray}{rCl}
\operatorname{Tr}(C')^{-1} \le \operatorname{Tr}C^{-1}.
\end{IEEEeqnarray}

The eigenvalues of $C'$ coincide exactly with the eigenvalues of a matrix constructed as a convex combination of matrices with equal observation times for each $k$, that is:
\begin{IEEEeqnarray}{rCl}
\operatorname{Tr}C&=&N \sum \alpha_k k\\
\lambda_{\text{max}}^{(C)}&=& \sum \alpha_k k^2
\end{IEEEeqnarray}
and the $(N-1)$ smaller eigenvalues are equal.

As was already shown in Proposition 2, $\operatorname{Tr}C^{-1}$ is minimized over all such matrices when there is only one $k$ for which $\alpha_k \ne 0,$ that $k$ being $\frac{N}{2}$ ($N$ even) or $\frac{N+1}{2}$ ($N$ odd). Hence this $C$ is the global minimizer overall $C$ of the form
\begin{IEEEeqnarray}{rCl}
C=B^\top T B
\end{IEEEeqnarray}
\end{document}